\documentclass[iop,twocolappendix]{emulateapj}

\usepackage{animate}
\usepackage{multirow}

\shorttitle{Radio Models. I.}
\shortauthors{Ribeiro et al.}

\begin{document}


\title{Radio Frequency Models of Novae in eruption. I. \\ The Free-Free Process in Bipolar Morphologies}

\author{V.~A.~R.~M.~Ribeiro,\altaffilmark{1,9} L.~Chomiuk,\altaffilmark{2,3} U.~Munari,\altaffilmark{4} W.~Steffen,\altaffilmark{5} N.~Koning,\altaffilmark{6} T.~J.~O'Brien,\altaffilmark{7} T.~Simon,\altaffilmark{1} P.~A.~Woudt,\altaffilmark{1} and M.~F.~Bode\altaffilmark{8}}


\altaffiltext{1}{Astrophysics, Cosmology and Gravity Centre, Department of Astronomy, University of Cape Town, Private Bag X3, Rondebosch 7701, South Africa}\email{vribeiro@ast.uct.ac.za}
\altaffiltext{2}{Department of Physics and Astronomy, Michigan State University, East Lansing, Michigan 48824, USA}
\altaffiltext{3}{National Radio Astronomy Observatory, 520 Edgemont Rd, Charlottesville, VA, USA}
\altaffiltext{4}{NAF Astronomical Observatory of Padova, I-36012 Asiago (VI), Italy}
\altaffiltext{5}{Instituto de Astronom\'ia, Universidad Nacional Aut\'onoma de M\'exico, C.P. 22860, Ensenada, Mexico}
\altaffiltext{6}{Department of Physics \& Astronomy, University of Calgary, Calgary, Alberta T2N 1N4, Canada}
\altaffiltext{7}{Jodrell Bank Centre for Astrophysics, University of Manchester, Manchester M13 9PL, UK}
\altaffiltext{8}{Astrophysics Research Institute, Liverpool John Moores University, IC2 Liverpool Science Park, 146 Brownlow Hill, L3 5RF, UK}
\altaffiltext{9}{South African Square Kilometer Array Fellow.}


\begin{abstract}
Observations of novae at radio frequencies provide us with a measure of the total ejected mass, density profile and kinetic energy of a nova eruption. The radio emission is typically well characterized by the free-free emission process. Most models to date have assumed spherical symmetry for the eruption, although it has been known for as long as there have been radio observations of these systems, that spherical eruptions are to simplistic a geometry. In this paper, we build bipolar models of the nova eruption, assuming the free-free process, and show the effects of varying different parameters on the radio light curves. The parameters considered include the ratio of the minor- to major-axis, the inclination angle and shell thickness (further parameters are provided in the appendix). We also show the uncertainty introduced when fitting spherical model synthetic light curves to bipolar model synthetic light curves. We find that the optically thick phase rises with the same power law ($S_{\nu} \propto t^2$) for both the spherical and bipolar models. In the bipolar case there is a ``plateau'' phase -- depending on the thickness of the shell as well as the ratio of the minor- to major-axis -- before the final decline, that follows the same power law ($S_{\nu} \propto t^{-3}$) as in the spherical case. Finally, fitting spherical models to the bipolar model synthetic light curves requires, in the worst case scenario, doubling the ejected mass, more than halving the electron temperature and reducing the shell thickness by nearly a factor of 10. This implies that in some systems we have been over predicting the ejected masses and under predicting the electron temperature of the ejecta.
\end{abstract}

\keywords{(stars:) novae, cataclysmic variables --- radio continuum: stars}

\section{Introduction}
A nova eruption occurs in a binary system following extensive accretion of hydrogen rich material on to the surface of a white dwarf primary from a less evolved secondary star. The eruption is well established to be a thermonuclear runaway on the surface of the white dwarf \cite[see, e.g.,][]{SIH08}. The eruption ejects of order 10$^{-7}$ to 10$^{-3}$ M$_{\sun}$ of matter at velocities of order hundreds to thousands of kilometers per second \citep[e.g.,][]{BE08,B10}. Since the white dwarf is not destroyed in the explosion, it may accrete more matter from the secondary star, either a main sequence star, sub-giant or giant star, and go into a cycle of eruptions. These eruptions can recur on time scales of years to millions of years, governed by properties the white dwarf, and the accretion \citep{SST85,TL86,YPS05}. Therefore, depending on the details of the white dwarf, including its composition and mass, plus accretion rate and ejected mass, we may expect the white dwarf to grow in mass, or not. If the mass of the white dwarf does grow \citep[e.g.,][]{NST13}, it may reach the Chandrasekhar limit and end in an accretion induced collapse to form a neutron star \citep[in the case of an ONe white dwarf; see, e.g.,][]{RGI96}, or grow and explode as a Type Ia supernova \citep[in the case of a CO white dwarf; for an extensive review see][]{dSOM13}.

Novae are now known to emit at all wavelengths, from $\gamma$--rays to radio; each providing vital information about the system parameters at the onset of the eruption. For example, observations at radio frequencies are of particular interest due to the fact that we can derive global properties of the ejecta, since radio emission is dominated by simple thermal free-free emission and does not suffer from interstellar extinction. Radio light curves, therefore, provide us with a measure of the total ejected mass, density profiles, and kinetic energy \citep{SB08,H96} and the distance once ejection velocity is known.
\begin{figure*}
\centering
\includegraphics[width=0.8\textwidth]{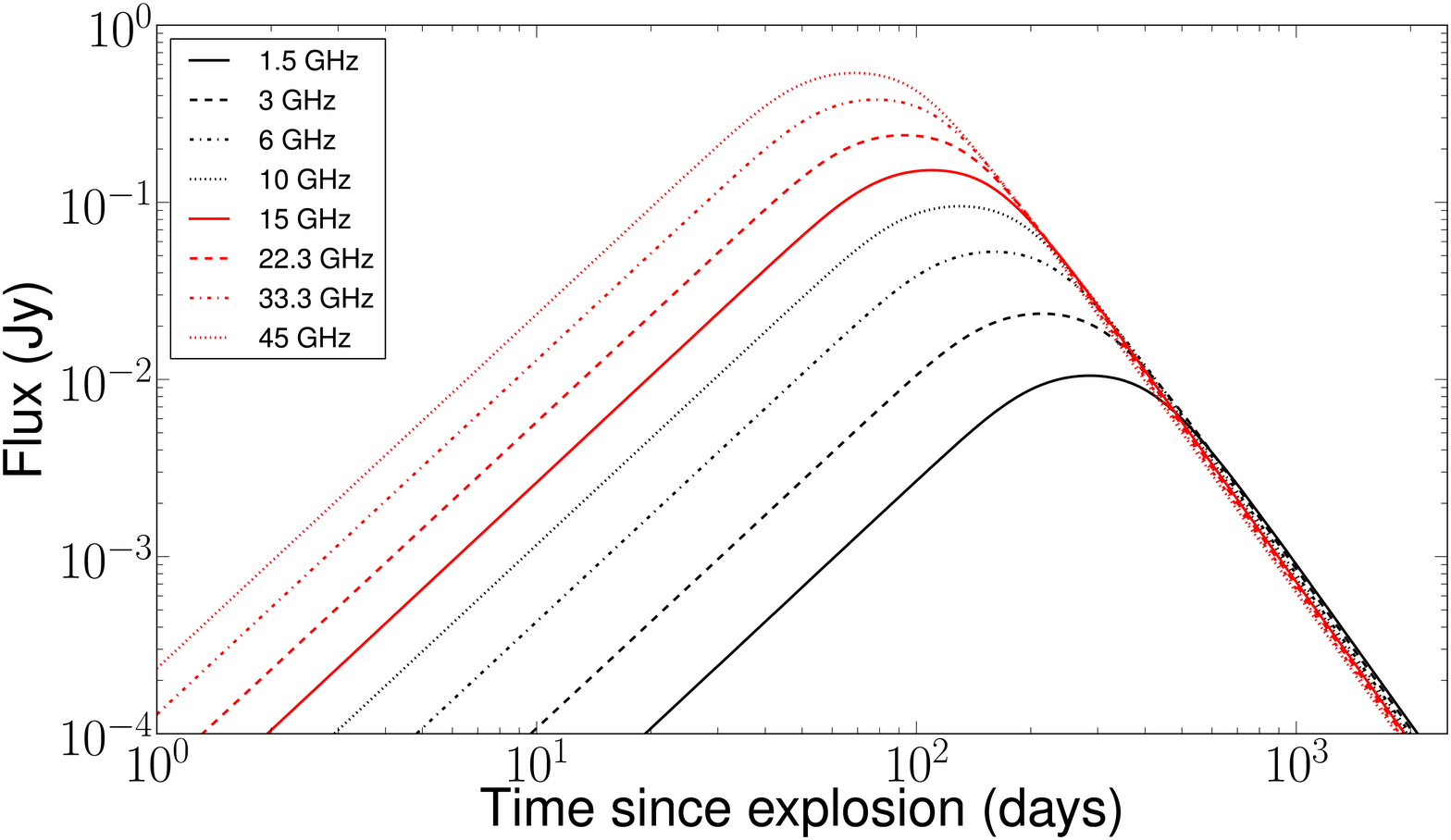}
\includegraphics[width=0.8\textwidth]{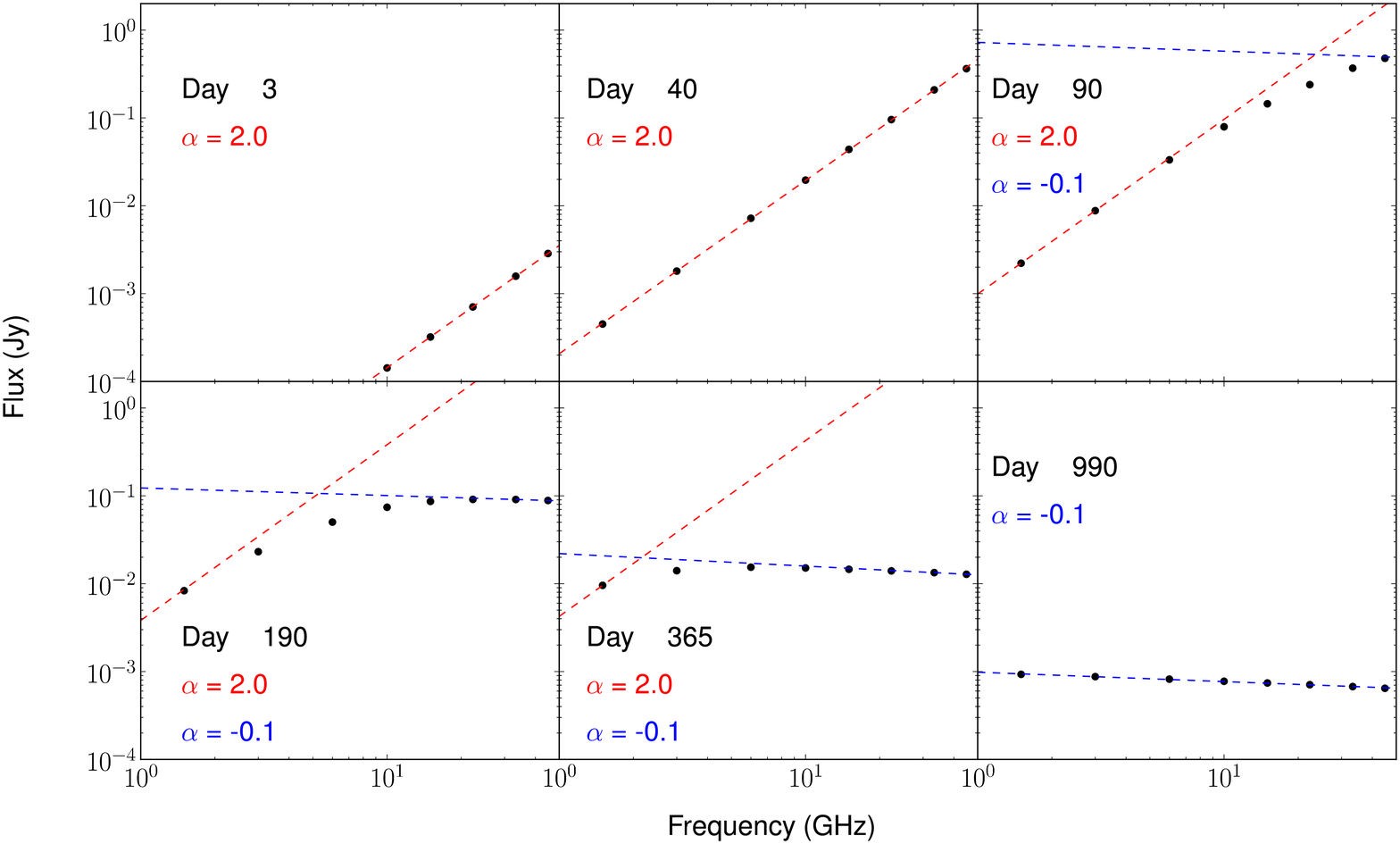}
\caption{$Top$ -- Synthetic radio light curves for a {\it spherical} eruption at a distance of 1~kpc, $M_{\rm ej} =~1~\times~10^{-4}$~M$_\sun$, $V_{\textrm{max}}$~=~3000~km~s$^{-1}$, $T_{\rm e}$ = 17000 K, and the ratio of the inner to outer radius of the shell of 0.25. The different frequency colors and line styles are as in the labels and Table~\ref{tb:band}. $Bottom$ -- spherical model spectral evolution at different dates demonstrating how the spectral index changes from completely optically thick ($\alpha$~=~2.0) to thin ($\alpha$~=~$-$0.1). Furthermore, to guide the eye lines are drawn to show the evolution of the spectral index from completely optically thick to thin. (A color version of this figure is available in the online journal.)}
\label{sphere}
\end{figure*}

The first radio light curves of novae \citep[HR~Del, FH~Ser, and V1500~Cyg;][]{HW70,WH71} were interpreted in terms of spherically symmetric ejecta with emission arising from the free-free process. The observed radio light curves, with spherical ejecta or otherwise, were described to arise from either a finite shell with an homologous expansion \citep{SP77,HWV79}, or a wind with a  constant velocity and mass loss rate \citep{K83}. The typical model consists of a $1/r^2$ density profile, with a constant temperature with time and radius in the ejecta, where $T\sim$~10$^4$~K.

At the beginning of the eruption, the spherically symmetric shell is optically thick, at all frequencies, increasing in flux density (proportional to the surface area of the shell, as seen on the plane of the sky) and follow the Planck function, hence a spectral index $\alpha$~=~2.0 (where $S_{\nu} \propto \nu^{\alpha}$, is the flux density and $\nu$ is the observed frequency). The flux density at this stage depends on the distance to the nova, electron temperature and the expansion velocity of the radio photosphere. As the ejecta expand, the density drops and the radio photosphere begins to recede, becoming optically thin at higher frequencies first. The flux density eventually peaks and starts to turn over at a particular frequency, when the photosphere starts to recede. The time when the peak and turn over happen depends on the ejected mass, density profile and electron temperature \citep[see Fig.~\ref{sphere};][]{HWV79,SB08}. When the ejecta are completely optically thin, at a given frequency, the spectral index is flatter, and ultimately $\alpha$~=~$-$0.1.

Observations of novae in the radio, over the last two decades, have shown that our earlier assumption of a spherically symmetric expanding shell may not always be the case \citep[see, e.g.,][and references therein for extensive reviews on the subject]{SP77,H96,SB08,RCS12}. \citet{SP77} noted that solely looking at the radio light curve does not permit us to distinguish between a spherical shell and a polar shell (which are just portions of a spherical shell), in particular to determine cone angles, and orientation of the polar shell. However, \citet{HO07} applied spherically symmetric and ellipsoidal models to the eruption of V723~Cas which was imaged with MERLIN and could not find differences between these two models with a simulated 12-h track, while on a 24-h track the ellipsoidal shell model could be retrieved during the optically thick phase. As the shell becomes optically thin, the images detect only the brighter emission coming from the inner shell boundary. Historically, there have been no clear signatures of asymmetries from the radio light curves. Therefore to break this degeneracy in determining the geometry from the radio light curve, complementary imaging is required \citep{OBP06,SRM08}.

In the optical, nova ejecta have been resolved to show a myriad of structures far from spherical; these include bipolar morphologies, prolate structures with equatorial and tropical rings \citep[e.g.,][]{H72,S83,SOD95,GO98,GO00,B02,KWS02,HO03,RBD09,WSK09}. Furthermore, nebular line profiles are well replicated with bipolar geometries \citep[e.g.,][]{H72,GO99,RMV13,SSA13,SAS13}.

In this paper, we demonstrate the effects of bipolar shells on the radio light curves during a nova eruption. As commonly assumed in novae, expansion is into a vacuum and no intervening interstellar material is present, such as that expected from systems with strong winds \citep[e.g.,][]{OBP06,CKR12}. Furthermore, we do not account for any other complicated morphologies as observed in, for example, V2672~Oph, where there was a combination of a prolate structure -- where the density appeared to decline faster -- and polar and equatorial rings \citep{MRB11}. In keeping with previous literature, at radio frequencies, we have kept the assumption that the filling factor is unity and there is no clumpyness -- we leave this as a discussion point later -- as well as that we assume an instantaneous ejection. In section~\ref{model}, we describe our modeling procedures, starting from a spherical symmetric shell and then changing this to a bipolar shell. In section~\ref{results}, we present the results of this change, and show the effect of varying different parameters on the radio light curve. Finally, in section~\ref{disc}, we discuss the relevance of our results, and provide conclusions and future work.

\section{Modeling Procedure}\label{model}
We aim to investigate the effect of bipolar and non-homogeneous  structure of the ejecta on their spatially-unresolved radio emission. To this aim we construct a complex geometry of a nova ejecta interactively in a 3D interface, within {\sc shape}\footnote{Available from \url{http://www.astrosen.unam.mx/shape}} \citep[Version 5;][]{SKW11}. The structure is then transferred to a 3D grid on which radiation transfer is computed via ray tracing to the observer. Radiation transfer is based on emission and absorption coefficients which are provided as a function of physical parameters such as density, temperature and wavelength. As the rays emerge from the grid, images and spectra are generated. Temporal evolution is simulated when a model of the structure's expansion is provided. The time sequence of the output is then generated automatically. The emissivity, used to generate the synthetic images, is proportional to the density squared.

In the Physics module within {\sc shape}, we input the free-free emission ($\epsilon_{\nu}$ in units of W~m$^{-3}$~sr$^{-1}$~Hz~$^{-1}$) and absorption ($\kappa_{\nu}$ in units of m$^{-1}$) coefficients at a given frequency ($\nu$ in Hz), as \citep{BS09}:
\begin{equation}
\epsilon_{\nu} = 6.8 \times 10^{-51} Z^2 T_{\rm e}^{-0.5} N_{\rm e} N_{\rm z} \bar{g}_{ff}(\nu,T_{\rm e}) \exp{-\frac{h\nu}{kT_{\rm e}}} ,
\label{eq1}
\end{equation}
\begin{equation}
\kappa_{\nu} (T_{\rm e}) =  1.77 \times 10^{-12} T_{\rm e}^{-1.5} Z^2 N_{\rm e} N_{\rm z} \nu^{-2} \bar{g}_{ff}(\nu,T_{\rm e}),
\label{eq2}
\end{equation}
respectively, where, $N_{\rm e}$ = $N_{\rm z}$ are the electron and ion mass densities, $Z$ is the atomic number ($Z$~=~1 for a singly ionised atom) and $T_{\rm e}$ the electron temperature. All values are in SI units. The Gaunt factor, $\bar{g}_{ff}(\nu,T_{\rm e})$,  in the Rayleigh Jeans approximation, $h \nu \ll kT_{\rm e}$, has only a logarithmic dependence on frequency \citep{B66}:
\begin{equation}
\bar{g}_{ff}(\nu,T_{\rm e}) = \frac{\sqrt{3}}{\pi} \left [ 17.7 + \ln \frac{T_e^{1.5}}{\nu} \right ].
\end{equation}

\subsection{Spherical Models}
We first demonstrate that we can reproduce the classical spherical models within {\sc shape}. Our spherical model has a shell thickness of 0.25, defined as the ratio of the inner to the outer radius of the shell. We define the inner radius to be $0.25 \times t \times V_{\textrm{max}}$, where $t$ is time since eruption and $V_{\textrm{max}}$ the maximum velocity; conversely, the outer radius is $t \times V_{\textrm{max}}$. This assumes a velocity linearly proportional to the radius. The input parameters are $V_{\textrm{max}}$~=~3000~km~s$^{-1}$, the electron temperature, $T_{\rm e}$~=~17000~K, ejected mass, $M_{\textrm{ej}}$~=~1$\times$10$^{-4}$~M$_{\sun}$, a 1/$r^2$ density distribution and a distance of 1~kpc. These values were chosen primarily from radio observations \citep[e.g.,][]{HWV79,H96,THS88,HOE05}.

Monte Carlo line profile modelling of the structure of the nova ejecta assume a 1/$r^3$ density profile \citep{SAS13,SSA13} which is also used in photo-ionisation models \citep[e.g.,][]{SSS01,SSB03,VSS05,MSD11} while, for example, \citet{MSH08} could not find a good fit using values for the exponent of 0, $-1$, and $-3$ with 1/$r^2$ providing a better fit, and morpho-kinematical modelling of the [O~{\sc iii}] 4959/5007~\AA\ emission line by \citet{RMV13} assumed a constant density distribution. The photo-ionisation models above are based on {\sc cloudy} \citep{FKV98} which are in 1D. The full 3D treatment can be achieved, for example with {\sc moccasin} \citep{EBS03} however, these are computationally intensive in order to explore the full parameter space. Pseudo-3D models based on {\sc cloudy} are also being developed \citep[{\sc rainy3d},][]{MD09,MD11} which can also account for clumpyness. We also note that a shell thickness of 0.25 is higher than that derived from photo-ionisation modelling \citep[e.g.,][]{VSS05,MSH08,MSD11} -- although photo-ionisation modelling should also be constrained with multifrequency and multi-epoch observations \citep[e.g.,][]{SSS01,S14} -- and indeed also from recent geometrical studies of the resolved ejecta of GK Per \citep[][although this object may be somewhat of a special case]{LCS12}.

The frequencies explored are targeted towards observational bands of the Karl G. Jansky Very Large Array and are given in Table~\ref{tb:band} and the results are presented in Fig.~\ref{sphere}. We compared the spherical models produced here with the numerical integration of a spherical shell modelled after \citet{HWV79} and \citet{HOE05}. To demonstrate that the models developed in this paper are equivalent to those previously published in the first row in Table~\ref{tb2} we show the fit of one such model to a spherical model from this paper.
\begin{table}
\caption{Bandwidths applied to the models, based on those for the Karl G. Jansky Very Large Array.}
\centering
\begin{tabular}{lccl}
Band & Range (GHz) & Centre (GHz) & Colour/linestlye\\
\hline\hline
20 cm (L) 	& \phn1.0--\phn2.0 & \phn1.5 & Black/solid \\
13 cm (S) & \phn2.0--\phn4.0 & \phn3.0 & Black/dashed \\
6 cm (C) & \phn4.0--\phn8.0  & \phn6.0 & Black/dashdot \\
3 cm (X) & \phn8.0--12.0  & 10.0 & Black/dotted \\
2 cm (Ku) & 12.0--18.0  & 15.0 & Red/solid \\
1.3 cm (K) & 18.0--26.5 & 22.3 & Red/dashed \\
1 cm (Ka) & 26.5--40.0 & 33.3 & Red/dashdot \\
0.7 cm (Q) & 40.0--50.0 & 45.0 & Red/dotted \\ [0.4ex]
\hline
\end{tabular}
\label{tb:band}
\end{table}

\begin{table}
\caption{Best fit spherical model, at S- and Q-bands, for different values of the $squeeze$, assuming input models at a distance of 1~kpc, $M_{\rm ej}$~=~1$\times$10$^{-4}$~M$_{\sun}$, $T_{\rm e}$~=~17000~K, $V_{\textrm{max}}$~=~3000~km~s$^{-1}$, and a shell thickness of 0.25. When fitting the models we kept the distance and $V_{\textrm{max}}$ constant.}
\centering
\begin{tabular}{lccccc}
\multirow{2}{*}{$Squeeze$} & $i$ & $T_{\rm e}$ & $M_{\textrm{ej}}$ & \multirow{2}{*}{Shell} & Reduced $\chi^2$ \\
	& (degrees) & ($\times$10$^{4}$~K) &  ($\times$10$^{-4}$~M$_\sun$) & & (DOF=399) \\
\hline\hline
0.0 & -- & 1.73 & 0.98 & 0.24 & 0.47 \\
			\hline
\multirow{2}{*}{0.1} 	& 0 & 1.42 & 1.04 & 0.25 & 0.68 \\
			& 90	& 1.64 & 1.03 & 0.23 & 0.40 \\
			\hline
\multirow{2}{*}{0.2} 	& 0 & 1.17 & 1.11 & 0.26 & 1.73 \\
			& 90	& 1.54 & 1.09 & 0.22 & 0.42 \\
			\hline
\multirow{2}{*}{0.3} 	& 0 & 0.99 & 1.18 & 0.25 & 3.19 \\
			& 90	& 1.43 & 1.16 & 0.20 & 0.56 \\
			\hline
\multirow{2}{*}{0.4} 	& 0 & 0.84 & 1.27 & 0.24 & 4.54 \\
			& 90	& 1.33 & 1.23 & 0.18 & 0.95 \\
			\hline
\multirow{2}{*}{0.5} 	& 0 	& 0.72 & 1.37 & 0.21 & 6.07 \\
			& 90	& 1.23 & 1.32 & 0.16 & 1.73 \\
			\hline
\multirow{2}{*}{0.6} 	& 0 	& 0.63 & 1.47 & 0.17 & 8.27 \\
			& 90	& 1.12 & 1.42 & 0.13 & 3.52 \\
			\hline
\multirow{2}{*}{0.7} 	& 0 	& 0.55 & 1.62 & 0.13 & 14.46 \\
			& 90	& 1.01 & 1.57 & 0.10 & 7.14 \\
			\hline
\multirow{2}{*}{0.8} 	& 0 	& 0.48 & 1.81 & 0.08 & 36.24 \\
			& 90	& 0.90 & 1.77 & 0.06 & 22.46 \\
			\hline
\multirow{2}{*}{0.9} 	& 0 	& 0.42 & 2.18 & 0.04 & 128.37 \\
			& 90	& 0.78 & 2.17 & 0.03 & 58.01 \\
\hline
\end{tabular}
\label{tb2}
\end{table}

\subsection{Bipolar Models}
\begin{figure*}
\centering
\includegraphics[width=0.4\textwidth]{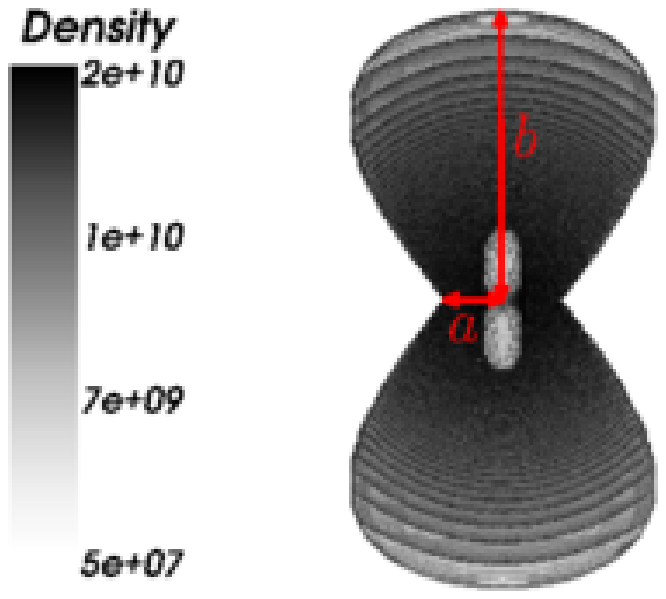}
\includegraphics[width=0.5\textwidth]{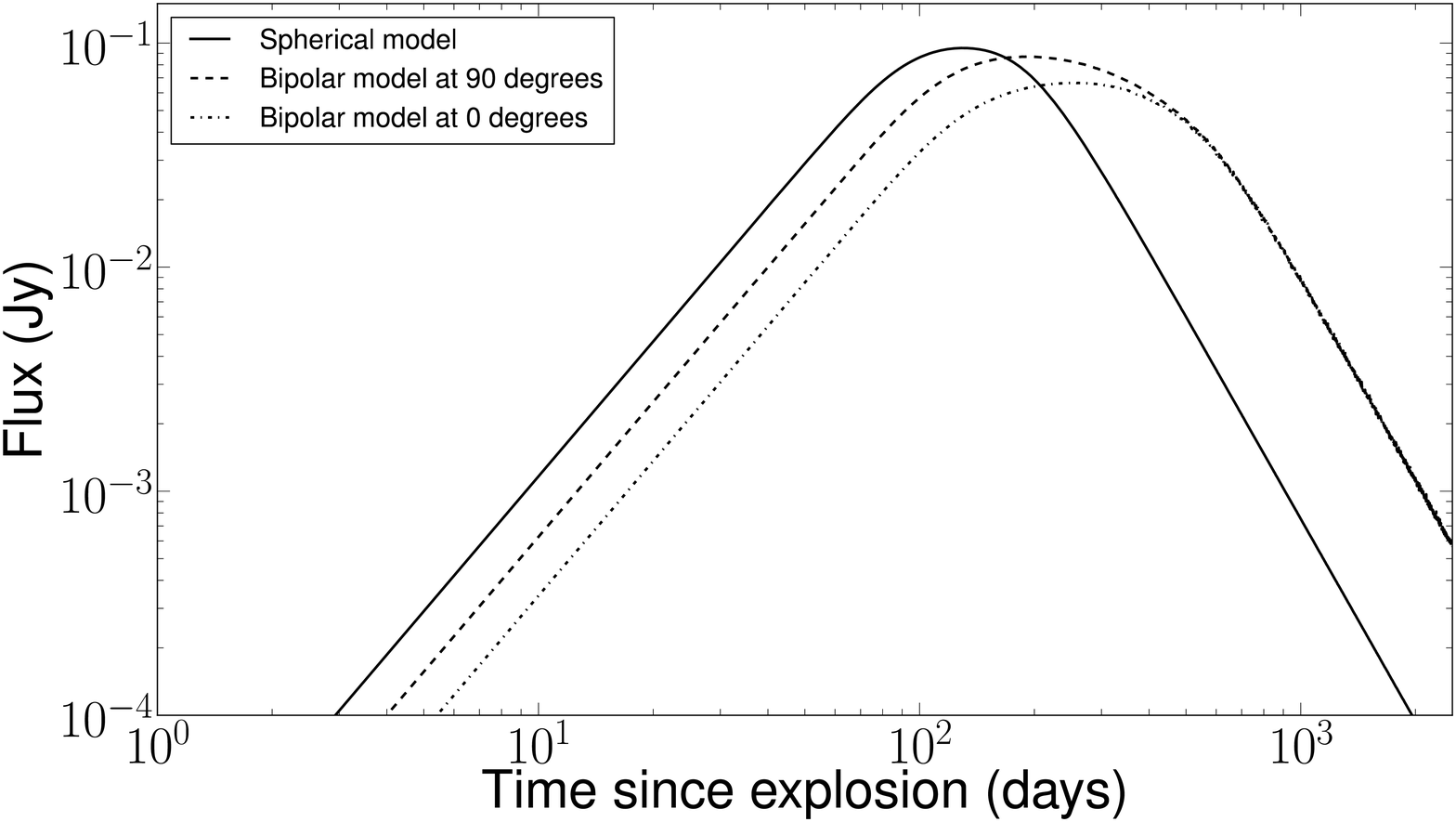}
\caption{$Left$ -- bipolar morphology used as input for the modeling, as seen at 90$^{\circ}$. The density is in electrons per cubic meter. The ratio of the semi-major to semi-minor axis ($a$ and $b$, respectively) is of 5 ($squeeze$ = 0.8) at 90$^{\circ}$ as viewed on the paper. $Right$ -- Comparison between the X-Band spherical model (solid black; as Fig.~\ref{sphere}) and a bipolar model, with the same initial parameters as the spherical model at 0 and 90$^{\circ}$ (black dashdot and dashed lines, respectively). (A color version of this figure is available in the online journal.)}
\label{bipolar}
\end{figure*}

Subsequently, we modify our spherical shell to a bipolar geometry, where the ratio of the major axis is 5 times greater than the minor axis (left hand panel, Fig.~\ref{bipolar}). All other system parameters are kept the same as in the spherical case. In this bipolar case, the maximum velocity along the major-axis is $V_{\textrm{max}}$~=~3000~km~s$^{-1}$, while in the minor axis is $V_{\textrm{minor}}$~=~600~km~s$^{-1}$, determined from the ratio of the axes. We use the $squeeze$ modifier to obtain the different axial ratios and is defined as $squeeze$ $= 1 - a/b$, where $a$ and $b$ are the semi-minor and -major axis, respectively.

\section{Results}\label{results}
\begin{figure*}
\plotone{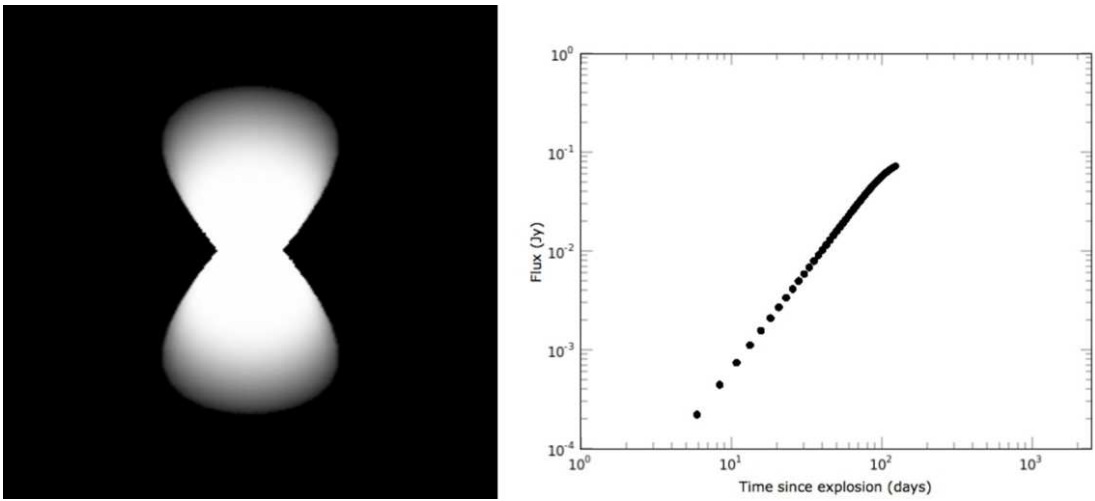}
\caption{Synthetic image and light curve (left and right panels, respectively) for the evolution of the bipolar nova shell, given in Fig.~\ref{bipolar}, as it expands from the optically thick phase to the optically thin decline, at an inclination of 90$^{\circ}$. (Animation available in the electronic version of the article.)}
\label{movie}
\end{figure*}

\begin{figure*}
\plottwo{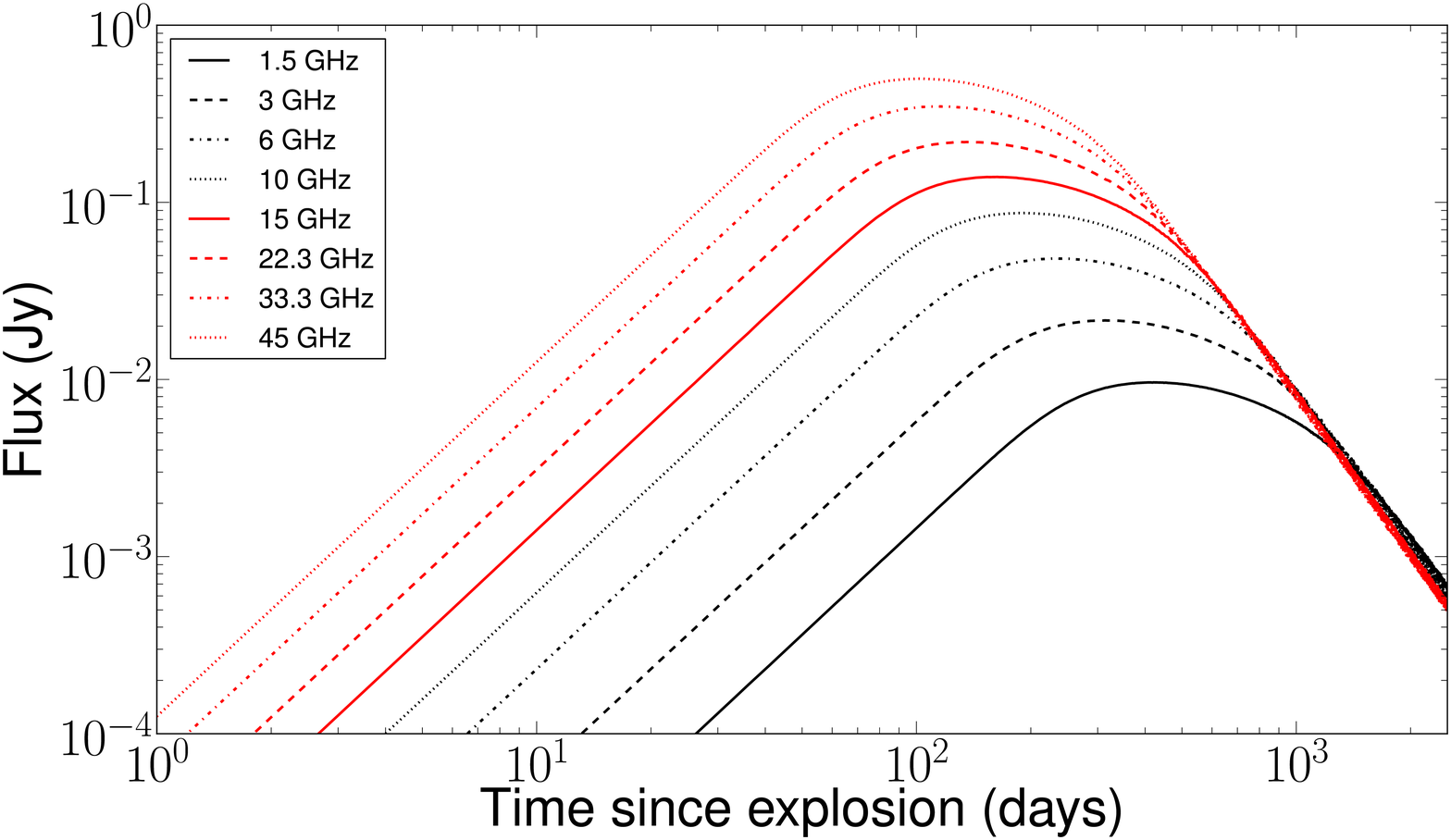}{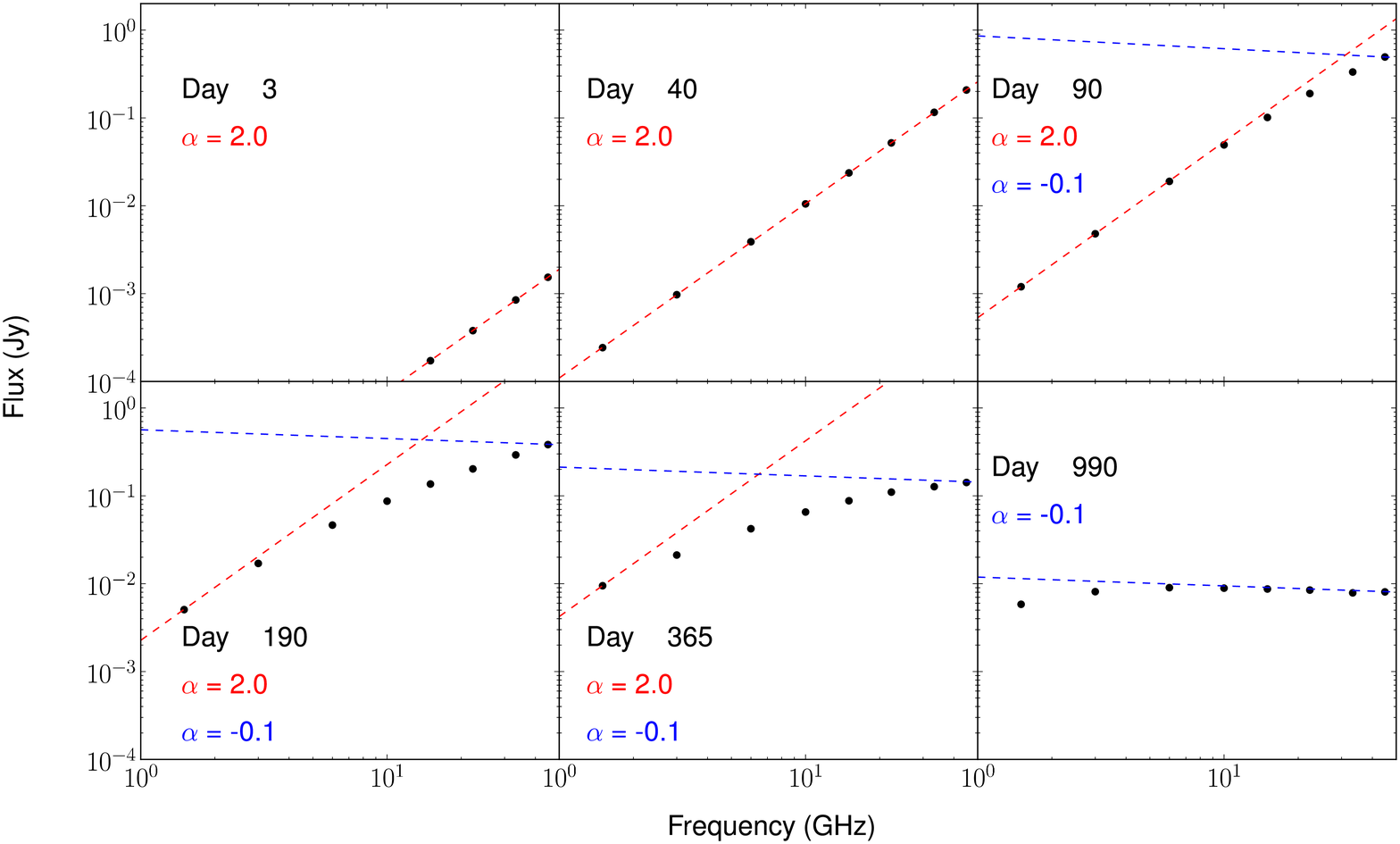}
\plottwo{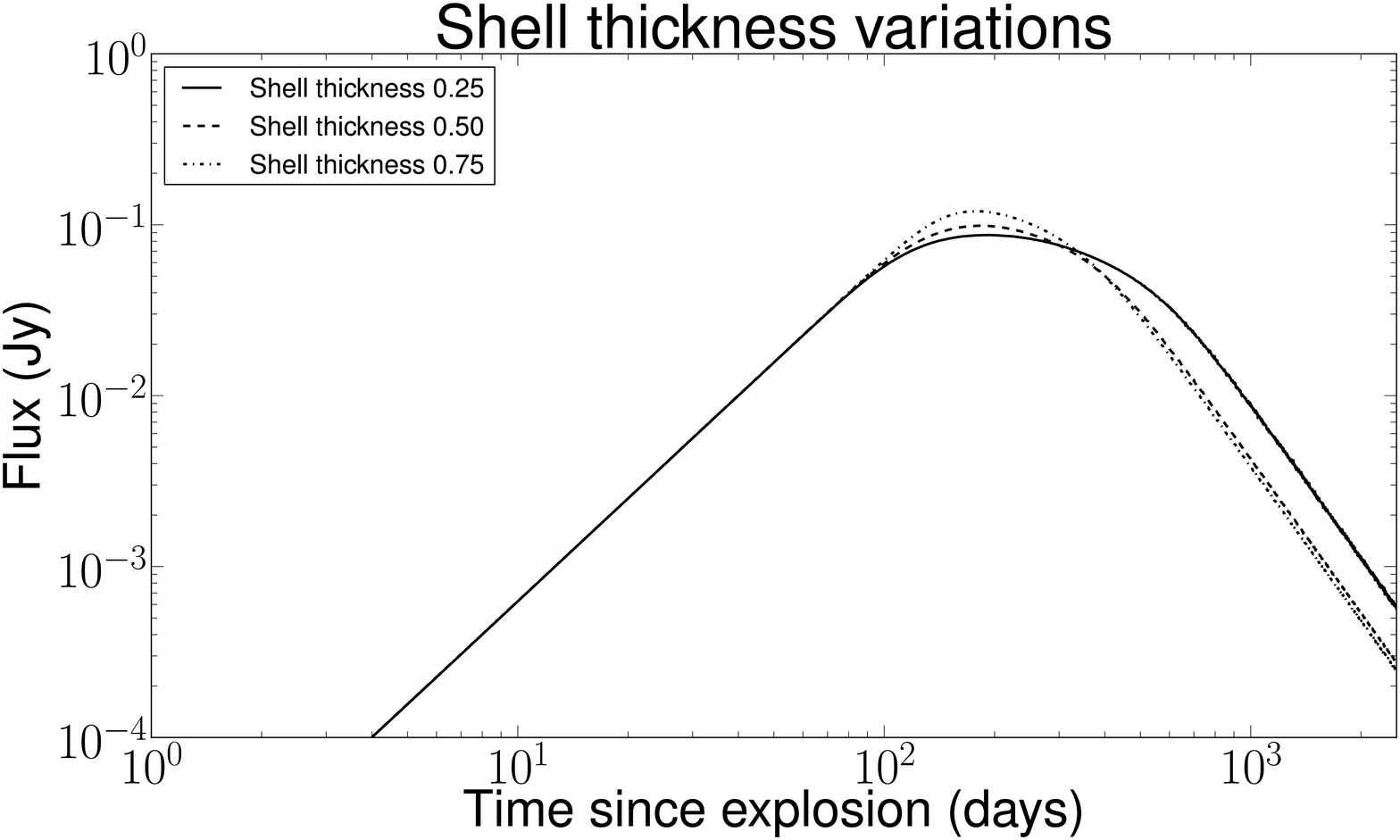}{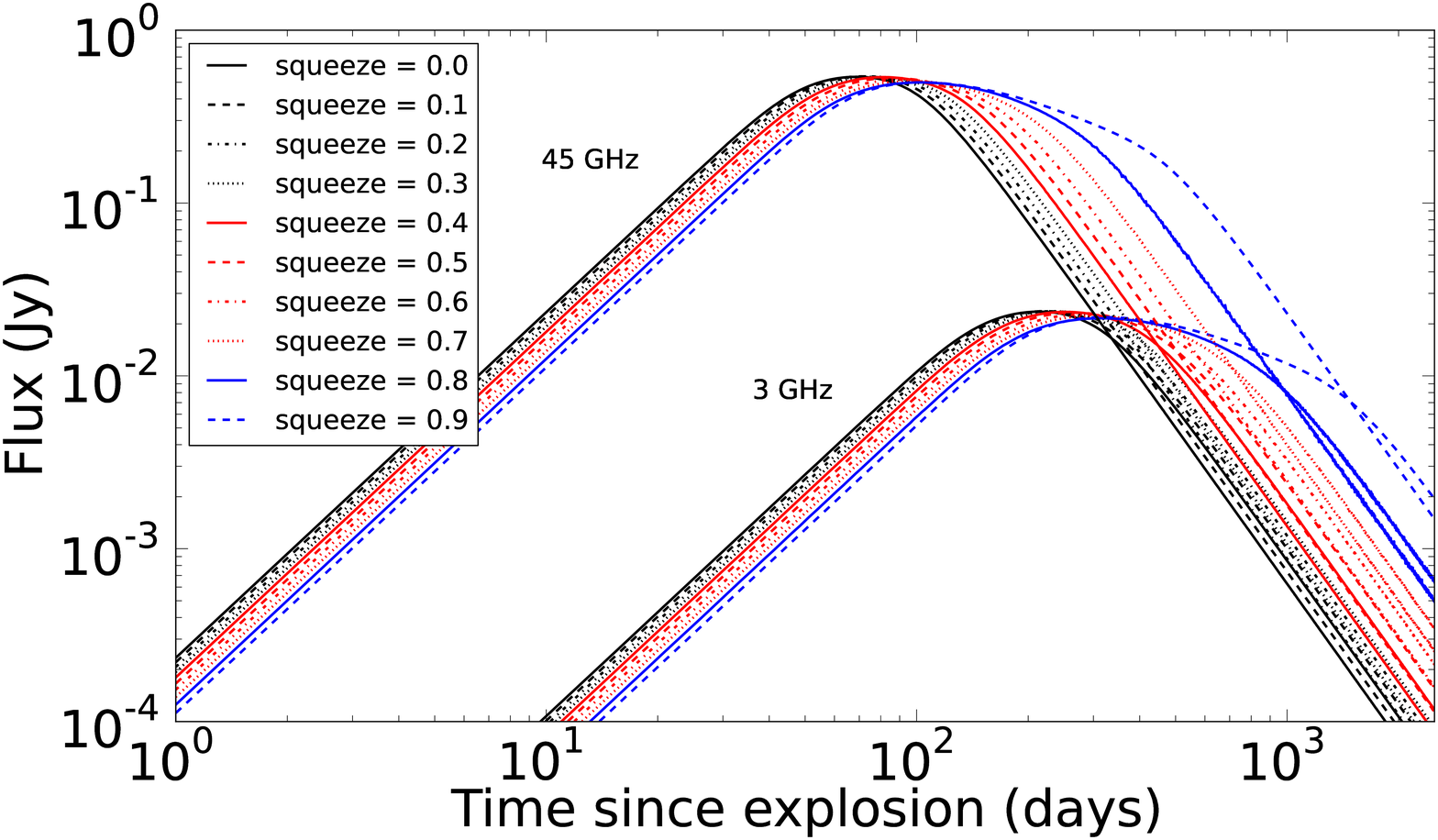}
\caption{Shown are the effects of changing different input parameters for a {\it bipolar} model with $M_{\textrm{ej}}$~=~1$\times$10$^{-4}$~M$_{\sun}$, $T_{\rm e}$~=~17000~K, $V_{\textrm{max}}$~=~3000~km~s$^{-1}$, and a shell thickness of 0.25. {\it Top left} -- are the bipolar model light curves at the frequencies defined in Table~\ref{tb:band} at an inclination of 90$^{\circ}$. {\it Top right} -- spectral index evolution at different dates. {\it Bottom left} -- the shell thickness is then modified to 0.5 and 0.75 (dashed and dash dot lines, respectively). In both cases to retrieve the same ejected mass the density is required to increase, hence the higher the peak. {\it Bottom right} -- varying the $squeeze$ value at two different frequencies (3~GHz and 45~GHz, lower and upper curves, respectively). It is noticeable that moving from a sphere ($squeeze$ = 0.0) to a bipolar morphology increases the ``plateau'' phase. (A color version of this figure is available in the online journal.)}
\label{bipolar_changes}
\end{figure*}

The right hand panel of Fig.~\ref{bipolar} shows the synthetic light curve for the bipolar model at two different inclinations (where an inclination $i$ = 90$^{\circ}$ corresponds to the orbital plane being edge-on, and $i$ = 0$^{\circ}$ being face-on) compared with the spherical model, at X-Band (10~GHz) and assuming the initial conditions as described in the previous section. Below we describe the evolution of the bipolar model synthetic light curve; and in Fig.~\ref{movie} we provide a time sequence of the evolution for a bipolar system at an inclination angle of 90$^{\circ}$ as a visual aid to the description:
\begin{itemize}
\item The initial optically thick rise phases are equivalent for both the spherical and bipolar models and follow the Planck function. However, the flux density is lower in the bipolar models due to the fact that it is proportional to the surface area of the shell (as viewed on the plane of the sky); in the bipolar model, depending on the inclination angle, only a certain fraction of the object is observed compared a spherical ejecta at the same phase of evolution. Furthermore, the flux density at this stage increases as $t^2$ in all three cases -- spherical and bipolar. Again, the spectral index at this time is $\alpha$~=~2.0 (upper right panel Fig.~\ref{bipolar_changes} and lower panel Fig.~\ref{sphere} for comparison).
\item For the same mass of ejecta, the bipolar model density will obviously be higher due to the fact that the volume is smaller. Depending on the inclination angle for the bipolar model, the peak flux density is around the same level or slightly below (90 and 0$^{\circ}$, respectively). The lower flux density, at 0$^{\circ}$, is due to the fact that the photospheres never reaches as large an area as if the ejecta where observed at, for example, 90$^{\circ}$. 
\item The light curve then enters a ``plateau'' phase while the photosphere recedes, which is dependent on the shell thickness and the $squeeze$ --  both reducing the length of the plateau for a decrease in the shell thickness and the degree of bipolarity (bottom panels in Fig.~\ref{bipolar_changes}) -- before entering the internal cavity in the ejecta and declining. The decline phase flux density is proportional to $t^{-3}$ and follows the same behavior as the spherical spectral index, $\alpha$~=~$-$0.1 (upper right panel, Fig.~\ref{bipolar_changes}), albeit at slightly higher flux density.
\end{itemize}

Furthermore, in Appendix~A we show the effects that changing various parameters have on the radio light curves. In all cases we start with the X-Band model and then change one parameter at a time. As in the spherical case, the higher the frequency the higher the flux densities and the light curve peaks earlier. Decreasing the ejected mass results in lower peak densities and earlier turnover as expected. Similarly, the higher the temperature the earlier the peak. Increasing the velocity causes the material to become optically thin earlier with higher flux density.

\begin{figure*}
\plotone{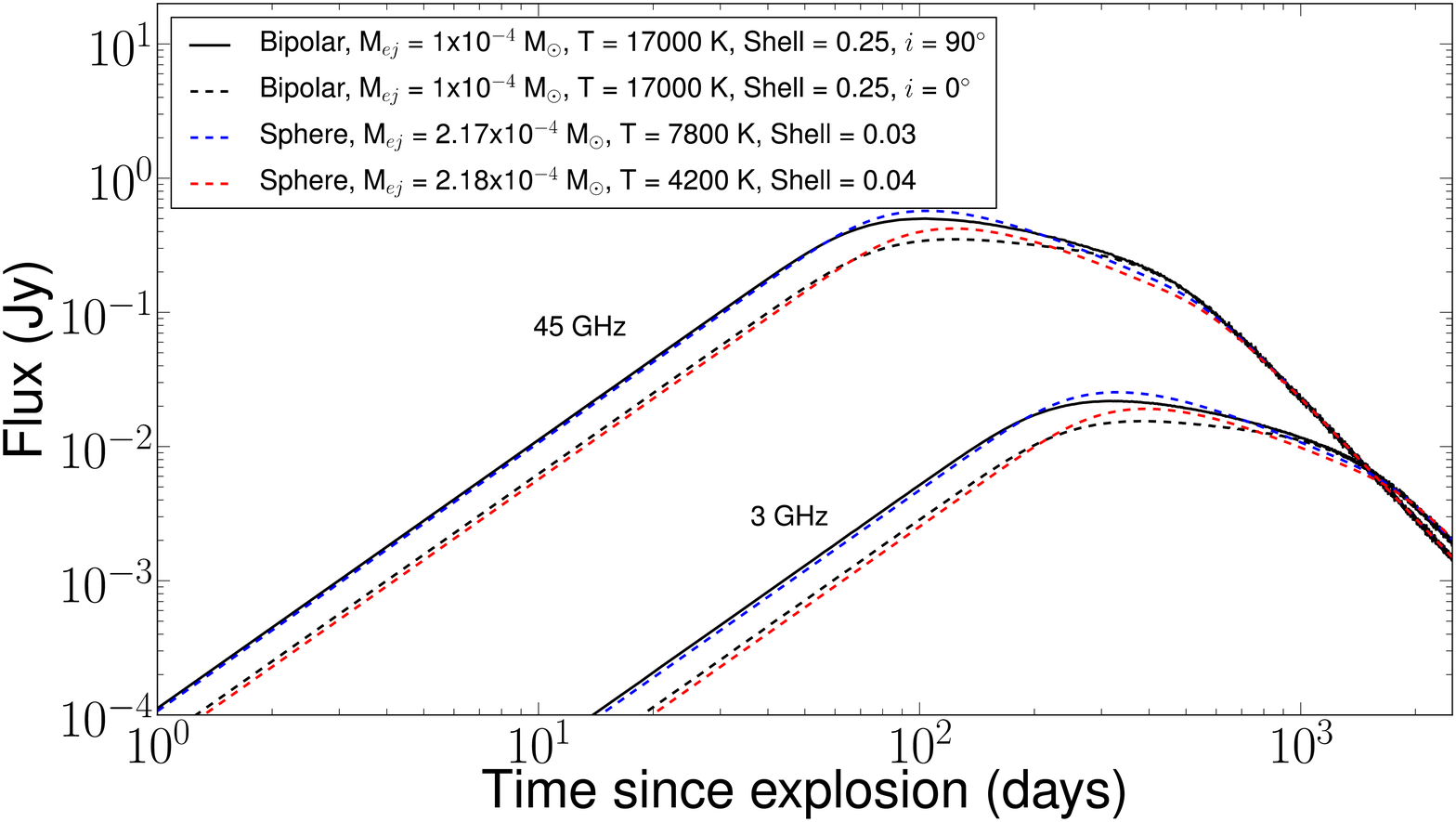}
\caption{Best fit spherical model synthetic light curves are fitted to bipolar model synthetic light curves. The input bipolar model, at 0 and 90 degrees (dashed and solid black lines, respectively), assumes $M_{\textrm{ej}}$~=~1$\times$10$^{-4}$~M$_{\sun}$, $T_{\rm e}$~=~17000~K, $V_{\textrm{max}}$~=~3000~km~s$^{-1}$, and a relative shell thickness of 0.25, at a distance of 1~kpc. The best fit spherical models are shown as blue and red dashed lines for the bipolar models at 90 and 0 degrees, respectively. The result shows that to find the best fit, keeping the maximum expansion velocity and distance constant, we require to double the ejected mass, increase the shell size (so that it reaches closer to the explosion site) and, depending on the inclination, reduce the temperature by more than half. (A color version of this figure is available in the online journal.)}
\label{bipolarvssphere}
\end{figure*}

Finally, we fit spherical model synthetic radio light curves to the bipolar model synthetic radio light curves at 0 and 90 degrees. We apply the same initial conditions as before, $M_{\textrm{ej}}$~=~1$\times$10$^{-4}$~M$_{\sun}$, $T_{\rm e}$~=~17000~K, $V_{\textrm{max}}$~=~3000~km~s$^{-1}$, and a shell thickness of 0.25, at a distance of 1~kpc. The results are shown in Table~\ref{tb2}, and in Fig.~\ref{bipolarvssphere} fits to the synthetic light curves for a $squeeze$ of 0.9 are shown. The general upshot of these results is that if we fit a spherical model to a bipolar model light curve we find an artificially high ejected mass, reduced temperature, and increased width of the shell, keeping the maximum velocity and distance the same. As illustrated by Table 2, the larger the departure from sphericity and the lower the inclination, the greater the difference. Furthermore, we show in Table~\ref{tb2} the results of the fit to a sphere ($squeeze$ = 0.0) to demonstrate the stability of the fitting.

\section{Discussion and conclusions}\label{disc}
First and foremost, the results presented above show some remarkable similarities in  the light curve between the different model morphologies. One notable difference however, is the shape of the light curve itself -- depending on the details of the shell thickness, and ratio of the minor- to major-axis. Disentangling the geometry and system parameters from the radio light curve is difficult, without further information from different wavelengths. If the light curve presents a longer ``plateau'' phase, as observed in Figs.~\ref{bipolar} and \ref{bipolar_changes}, we may assume that this is an indication of a bipolar morphology. As shown in the lower left panel Fig.~\ref{bipolar_changes}, we are able to reduce the ``plateau'' phase if we reduce the size of the shell. Therefore, all these factor will induce an error in the mass estimation hence it is imperative that we apply estimates of the ejecta geometry derived from optical line profiles or high-resolution imaging to the radio observations.

The results presented in Table~\ref{tb2} and Fig.~\ref{bipolarvssphere}, demonstrate that in some cases, in the literature, we may have overestimated the ejected masses by fitting spherical models to light curves that arise from a bipolar ejecta and underestimated the temperature of the ejecta. We require, therefore, that the geometry of the system is constrained early on after eruption \citep[see, e.g.,][]{RMV13,SSA13}. These results are a stepping stone towards solving the question of the order-of-magnitude discrepancy between the predicted and observed ejected masses (the observed being the higher masses; \citealt{PK95,GTW98,STW98,JCH99,G02,YPS05}). There are a number of issues that will affect the ejected masses; i.e. clumpyness, the filling factor, a realistic ejecta morphology (as described above the morphology is far from simplistic as a simple bipolar too), among others factors. The discrepancy appears to be predominantely in the fastest novae \citep[see, e.g.,][]{RCS12}.

These simple models of a bipolar morphology assuming the free-free emission process are, however, not sufficient to replicate the myriad of observed radio light curves. For example, V1723 Aql  presents a steep rise ($S_{\nu} \propto t^{3.3}$) during the optically thick phase \citep{KCR11,WSZ13}. Furthermore, the temporal and spectral evolution is different from that described in this paper. The radio light curves also show bumps \citep[e.g., V1324 Sco;][]{WSZ13}, where there is a phase where the flux density increases then falls, only subsequently to rise again before its final decline; kinematically we may understand this as arising from two distinct shells where the fastest moving shell becomes optically thin first and as the radio photosphere recedes towards the inner shell, that is still optically thick, the flux density rises again once the first shell becomes completely optically thin (we leave this as future work) -- this may  also explain some of the features observed in T~Pyx, for example as observed in \citet{NCR12}. However, the Russian doll structure described above does not account for the steep rise in the radio light curve, which may be due to a number of factors (for example, a variable temperature gradient in the ejecta left for future work). Indeed, current theoretical nova models do not predict a series of discrete, time separated mass ejections however, \citet{S13} has presented a model for the spectral and photometric evolution that does not require secondary ejection or winds. In terms of future observations, we require early, frequent temporal and spectral evolution of these sources with good enough time sampling. Such may be achieved with up-and-coming large radio surveys, such as ThunderKAT, on MeerKAT (a precursor telescope to the Square Kilometer Array). Furthermore, with improvements on very long baseline interferometry, we are able to resolve sources much earlier and with smaller angular scales than before which will give clues to the origin of they myriad of radio light curves.

In this paper we aim to show the effects bipolar models have on our understanding of radio nova light curves. We show the effects that changing various parameters have on the radio light curve and our main conclusion is that in some cases where spherical models have been fit to an eruption where bipolar geometries in fact are present, this may induce an error in overestimating the mass of a factor of 2. An immediate example is that of V703~Cas. \citet{HOE05} interpreted the light curve as arising from a spherical model and retrieved parameters, namely the mass and distance to the nova. The spherical model was later shown to be incorrect when \citet{LC09} concluded that the morphology of the ejecta was different in the different emission lines, and suggested different ejection events. \citet{LC09} also derived a revised distance to V723~Cas from the expansion parallax method suggesting the object was much closer than that derived from \citet{HOE05}. We are now building models to account for these changes to update the parameters of V723~Cas (Ribeiro et al., in prep.).

Lastly, in this first paper, we kept the radio models as simple and close to those already in the literature, at least at radio frequencies. A number of effects that were not consider but warrant some discussion include non-uniformity of the ejecta - both in terms of the filling factor and clumpyness as well as temperature variations in the ejecta. The ejecta, particularly at optical wavelengths, has been shown to be very clumpy \citep[e.g., HR~Del, GK~Per, RR~Pic, T~Pyx, AT~Cnc;][]{GO98,HO03,LCS12,SZW97,SMW12,SZM12,SOD95}. \citet{W94} had already suggested that the ejecta shell is not homogeneous, as measured from the optical line ratios of [O~{\sc i}] and that neutral gas could exist in clumps. The clumps may be formed from Rayleigh-Taylor instabilities \citep{LOB97} during the early phases, while at later stages, when the shell density decreases Kelvin-Helmholtz instabilities should occur \citep{CBE92,CJG11}. \citet{MD09,MD11} showed that the presence of clumps to a non-spherical shells can affect the mass determination by a factor of $\sim$5. Finally, the temperature has up to now been assumed to be constant through out the shell however, there is strong evidence that this is not always the case. \citet{MHV14} have modelled V1324~Sco, from a 1D model, in terms of shocks between a fast nova outflow and a dense external shell setting up a temperature gradient. In the \citet{MHV14} model, they account for shocks and ionization state of the medium, replicating with confidence the radio light curve.

\acknowledgments

VARMR would like to thank Tom Jarrett for providing computing facilities through the NRF SARChI, Brian Warner for reading an initial draft and discussions and Russ Taylor for useful discussions. VARMR acknowledges the South African SKA Project for funding the postdoctoral fellowship position at the University of Cape Town (UCT). TS acknowledges the National Society of Black Physicists, USA, for funding the PhD studentship at UCT. PAW kindly acknowledges support from UCT and the NRF. We are grateful to our referee, Steven Shore, for very constructive comments.

\appendix
\section{variations in mass, temperature and velocity for a bipolar model}
\begin{figure*}
\plottwo{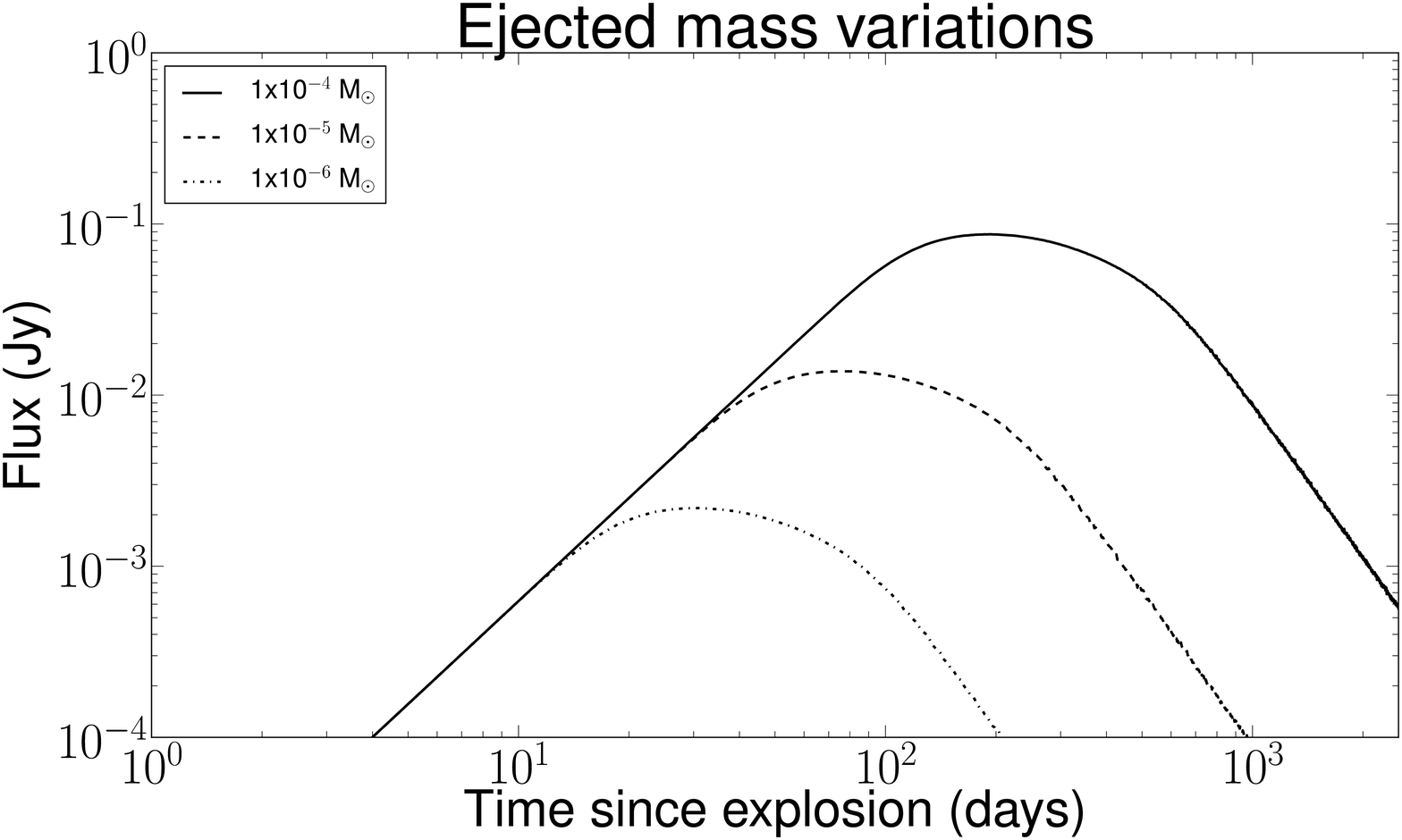}{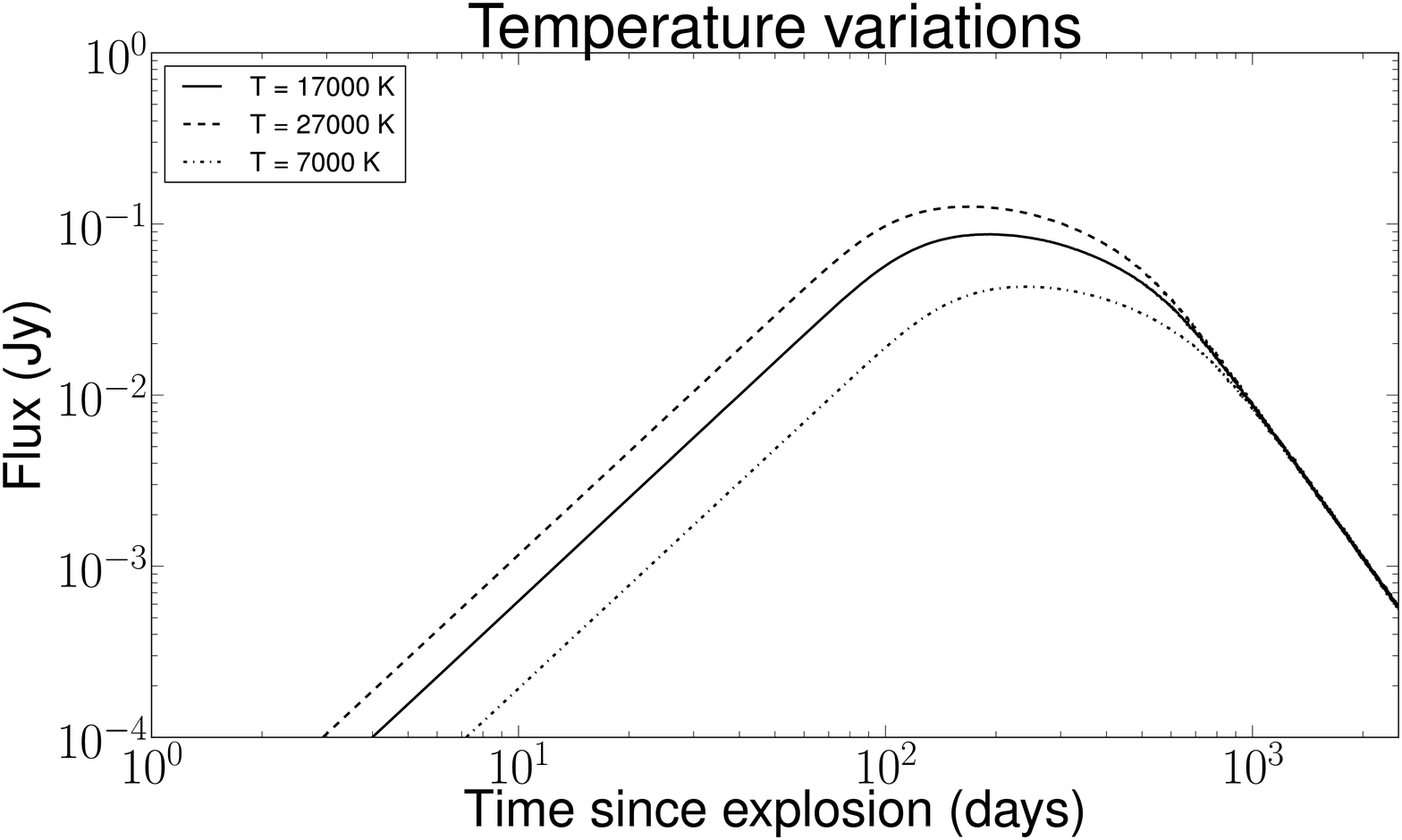}
\plotone{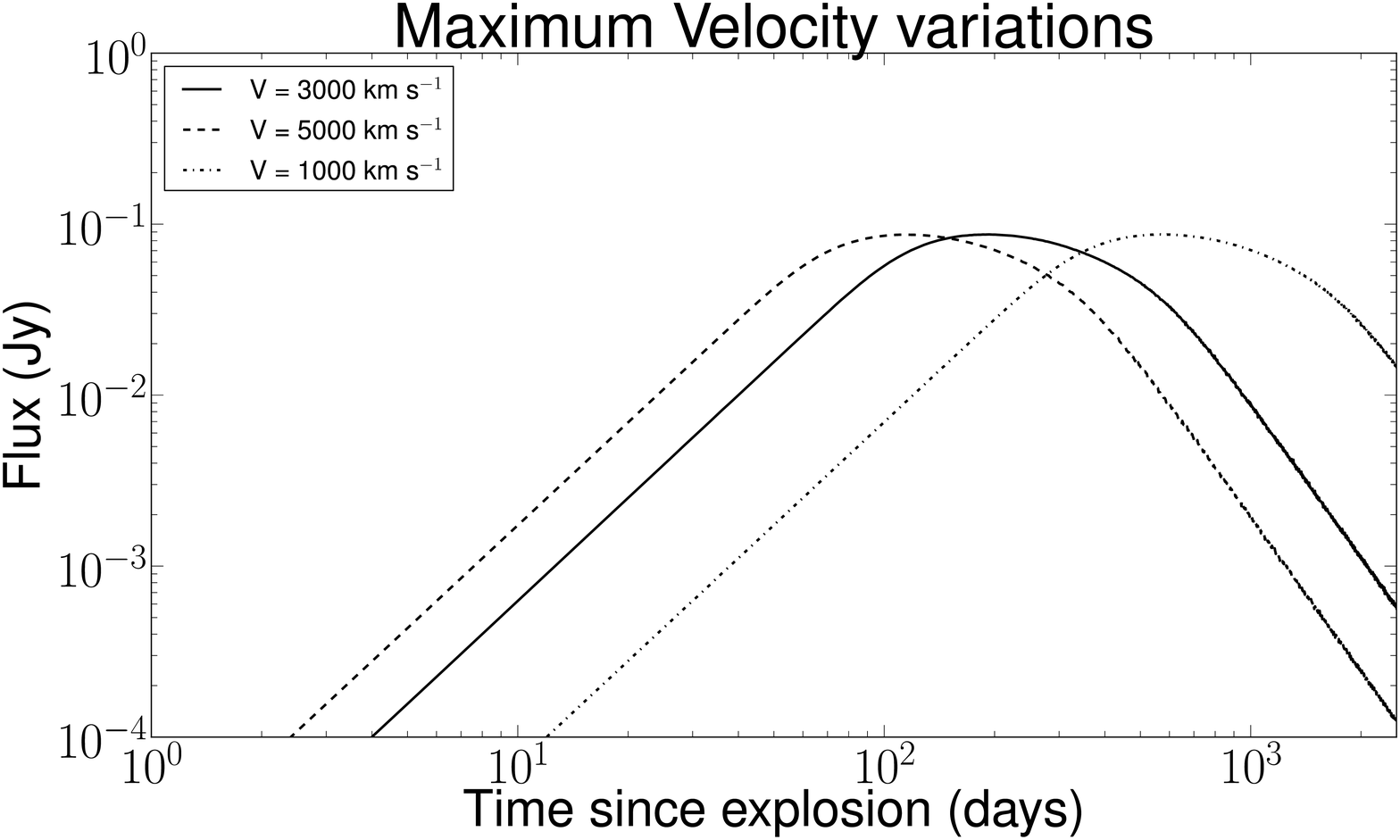}
\caption{The effects of changing different input parameters are shown for a {\it bipolar} model with $M_{\textrm{ej}}$~=~1$\times$10$^{-4}$~M$_{\sun}$, $T_{\rm e}$~=~17000~K, $V_{\textrm{max}}$~=~3000~km~s$^{-1}$, and a relative shell thickness of 0.25 (solid lines). {\it Top left} -- changing the ejected mass to 1$\times$10$^{-5}$~M$_{\sun}$ and 1$\times$10$^{-6}$~M$_{\sun}$ (dashed and dashdot lines, respectively) show a decrease in the peak flux density as well as an earlier turn over. {\it Top right} -- the temperature was varied to 27000 and 7000~K (dashed and dashdot lines, respectively). {\it Bottom} -- modifying the velocity to 5000 and 1000~km~s$^{-1}$ (dashed and dashdot lines, respectively) shifts the peak density to earlier or later times, respectively.}
\label{fig:bipolar_changes}
\end{figure*}

Fig.~\ref{fig:bipolar_changes} shows the effects of changing a number of input parameters. Changing the ejected mass, to higher values, will increase the flux density and cause a later peak/turnover as the material in the ejecta stays optically thick for longer. Increasing the temperature will shift the radio light curve to higher flux densities and a higher, and earlier, peak/turnover. While increasing the velocity will cause the radio light curve to shift to an earlier peak/turnover but at exactly the same peak flux density. These effects have exactly the same behaviour in a spherical model.


\end{document}